\documentclass[runningheads]{llncs}

\usepackage[utf8]{inputenc}
\usepackage[T1]{fontenc}

\usepackage{graphicx}
\usepackage[english]{babel}

\usepackage{listings}
\usepackage{algorithm}
\usepackage{xcolor}
\usepackage{amssymb}
\usepackage{mathtools}

\usepackage{subfig}

\usepackage{tikz}
\usepackage{tikzscale}

\usepackage[hidelinks]{hyperref}
\usepackage{cleveref}

\tikzstyle{every node}=[draw,fill=white,shape=circle, inner sep = 1pt]

\def\etal.{et\penalty50\ al.}

\def\ie,{i.e.,}
\def\ReDraw{\texttt{ReDraw}}

\let\cref\Cref
\let\lessdot\prec
\let\subset\subseteq

\newcommand{\R}{\mathbb{R}}

\begin{document}

\title{Force-Directed Layout of Order Diagrams using Dimensional Reduction}

\author{Dominik Dürrschnabel\inst{1,2}\orcidID{0000-0002-0855-4185}
  \and\\
  Gerd Stumme\inst{1,2}\orcidID{0000-0002-0570-7908}}

\institute{
  Knowledge \& Data Engineering Group,
  University of Kassel, Germany
  \and
  Interdisciplinary Research Center for Information System Design\\
  University of Kassel, Germany\\
  \email{\{duerrschnabel,stumme\}@cs.uni-kassel.de}}

\maketitle

\begin{abstract}
  Order diagrams allow human analysts to understand and analyze structural properties of ordered data.
  While an experienced expert can create easily readable order diagrams, the automatic generation of those  remains a hard task.
  In this work, we adapt force-directed approaches, which are known to generate aesthetically-pleasing drawings of graphs, to the realm of order diagrams.
  Our algorithm \ReDraw{} thereby embeds the order in a high dimension and then iteratively reduces the dimension until a two-dimensional drawing is achieved.
  To improve aesthetics, this reduction is equipped with two force-directed steps where one optimizes on distances of nodes and the other on distances of lines in order to satisfy a set of a priori fixed conditions.
  By respecting an invariant about the vertical position of the elements in each step of our algorithm we ensure that the resulting drawings satisfy all necessary properties of order diagrams.
  Finally, we present the results of a user study to demonstrate that our algorithm outperforms comparable approaches on drawings of lattices with a high degree of distributivity.

\end{abstract}

\setcounter{footnote}{0}

\section{Introduction}
\label{sec:introduction}

\emph{Order diagrams}, also called \emph{line diagrams}, \emph{Hasse diagrams} (or simply \emph{diagrams}) are a graphical tool to represent  ordered sets.
In the context of \emph{ordinal data analysis}, \ie, data analysis investigating ordered sets, they provide a way for a human reader to explore and analyze complex connections.
Every element of the ordered set is thereby visualized by a dot and two elements are connected by a straight line if one is lesser than the other and there is no element ``in between''.

The general structure of such an order diagram is therefore fixed by these conditions.
Nonetheless, finding good coordinates for the dots representing the elements such that the drawing is perceived as ``readable'' by humans is not a trivial task.
An experienced expert with enough practice can create such a drawing; however, this is a time-consuming and thus uneconomical task and therefore rather uncommon.
Still, the availability of such visualizations of order diagrams is an integral requirement for developing ordinal data science and semi-automated data exploration into a mature utensil; thus, making the automatic generation of such order diagrams an important problem.
An example of a research field that is especially dependent on the availability of such diagrams is  Formal Concept Analysis, a field that orders concepts derived from binary datasets in lattices.

The generated drawings have to satisfy a set of \emph{hard constraints} in order to guarantee that the drawing accurately represents the ordered set.
First of all, for comparable elements the greater element dot has to have a larger $y$-coordinate then the lesser element dot.
Secondly, no two element dots are allowed to be positioned on the same coordinates.
Finally, element dots are not allowed to touch non-adjacent lines.
Beside those hard criteria, there is a set of commonly accepted \emph{soft criteria} that are generally considered to make a drawing more readable, as noted in~\cite{Yevtushenko.2004}.
Those include maximizing the distances between element dots and lines, minimizing the number of crossing lines, maximizing the angles of crossing lines, minimizing the number of different edge directions or organizing the nodes in a limited number of layers.
While it is not obvious how to develop an algorithm that balances the soft criteria and simultaneously guarantees that the hard criteria are satisfied, such an algorithm might not even yield readable results as prior works such as~\cite{Demel.2018} suggests.
Furthermore, not every human reader might perceive the same aspects of an order diagram as ``readable''; conversely, it seems likely that every human perceives different aspects of a good drawing as important.
It is thus almost impossible to come up with a good fitness function for readable graphs.
Those reasons combined make the automatic generation of readable graph drawing - and even their evaluation - a surprisingly hard task.

There are some algorithms today that can produce readable drawings to some extent; however, none of them is able to compete with the drawings that are manually drawn by an expert.
In this paper we address this problem by proposing our new algorithm \ReDraw{} that adapts the force-directed approach of graph drawing to the realm of order diagram drawing.
Thereby, a physical simulation is performed in order to optimize a drawing by moving it to a state of minimal stress.
Thus, the algorithm proposed in this paper provides a way to compute sufficiently readable drawings of order diagrams.
We compare our approach to prior algorithms and show that our drawings are more readable under certain conditions or benefits from lesser computational costs.
We provide the source code\footnote{\url{https://github.com/domduerr/redraw}} so that other researchers do own experiments and extend it.

\section{Related Work}
\label{sec:related-work}

Order diagram drawing can be considered to be a special version of the graph drawing problem, where a graph is given as a set of vertices and a set of edges and a readable drawing of this graph is desired.
Thereby, each vertex is once again represented by a dot and two adjacent vertices are connected by a straight line.
The graph drawing problem suffers from a lot of the same challenges as order diagram drawing and thus a lot of algorithms that were developed for graph drawing can be adapted for diagram drawing.
For a graph it can be checked in linear time whether it is planar~\cite{Hopcroft.1974}), \ie, whether it has a drawing that has no crossing edges.
In this case a drawing only consisting of straight lines without bends or curves can always be computed~\cite[Sect. 4.2 \& 4.3]{Nishizeki.2004} and should thus be preferred.
For a directed graph with a unique maximum and minimum, like for example a lattice, it can be checked in linear time whether an upward planar drawing exists.
Then such a drawing can be computed in linear time~\cite[Sect.~6]{Battista.1999}.
The work~\cite[Sect.~3.2]{Battista.1999} provides an algorithm to compute straight-line drawings for ``serial parallel graphs'', which is a special family of planar, acyclic graphs.
As symmetries are often preferred by readers, the algorithm was extended~\cite{Hong.2000} to reflect them in the drawings based on the automorphishm group of the graph.
However, lattices that are derived from real world data using methods like Formal Concept Analysis rarely satisfy the planarity property~\cite{Albano.2017}.
The work of Sugiyama \etal.~\cite{Sugiyama.1981}, usually referred to as Sugiyama's framework, from 1981 introduces an algorithm to compute layered drawings of directed acyclic graphs and can thus be used for drawing order diagram.
Force-directed algorithms were introduced in~\cite{Eades.1984} and further refined in \cite{Fruchterman.1991}.
They are a class of graph drawing algorithms that are inspired by physical simulations of a system consisting of springs.

The most successful approaches for order diagram drawing are a work of Sugiyama \etal.~\cite{Sugiyama.1981} which is usually referred to as Sugiyama's framework and a work of Freese \cite{Freese.2004}.
Those algorithms both use the structure of the ordered set to decide on the height of the element dots; however, the approach choosing the horizontal coordinates of a vertex dot differ significantly.
While Sugiyama's framework minimizes the number of crossing lines between different vertical layers, Freese's layout adapts a force-directed algorithm to compute a three-dimensional drawing of the ordered set.
DimDraw~\cite{Duerrschnabel.2019} on the other hand is not an adapted graph drawing algorithm, but tries to emphasize the dimensional structure that is encapsulated in the ordered set itself.
Even though this approach is shown to outperform Freese's and Sugiyama's approach in~\cite{Duerrschnabel.2019}, it is not feasible for larger ordered sets because of its exponential nature.
in~\cite{Duerrschnabel.2019} proposes a method to draw order diagrams based on structural properties of the ordered set.
In doing so, two maximal differing linear extensions of the ordered set are computed.
The work in~\cite{Ganter.2004} emphasizes additive order diagrams of lattices.
Another force-directed  approach that is based on minimizing a ``conflict distance'' is suggested in~\cite{Zschalig.2007}.

In this work we propose the force-directed graph drawing algorithm \ReDraw{} that, similarly to Freese's approach, operates not only in two but in higher dimensions.
Compared to Freese's layout, our algorithm however starts in an arbitrarily high dimension and improves it then by reducing the number of dimensions in an iterative process.
Thus, it minimizes the probability to stop the algorithm early with a less pleasing drawing.
Furthermore, our approach gets rid of the ranking function to determinate the vertical position of the elements and instead uses the force-directed approach for the vertical position of dots as well.
We achieve this by defining a vertical invariant which is respected in each step of the algorithm.
This invariant guarantees that the resulting drawing will respect the hard condition of placing greater elements higher than lesser elements.

\section{Fundamentals and Basics}
\label{sec:fundamentals-basics}

In this section we recall fundamentals and lay the foundations to understand design choices of our algorithm.
This includes recalling mathematical notation and definitions as well as introducing the concept of force-directed graph drawing.

\subsection{Mathematical Notations and Definitions}
\label{sec:math-found}

We start by recalling some standard notations that are used throughout this work.
An \emph{ordered set} is a pair $(X,\leq)$ with ${\leq} \subseteq (X \times X)$ that is reflexive ($(a,a)\in{\leq}$ for all $a \in X$), antisymmetric (if $(a, b)\in{\leq}$ and $(b, a)\in{\leq}$, then $a = b$) and transitive (if $(a, b)\in{\leq}$ and $(b,c)\in{\leq}$, then $(a, c)\in{\leq}$).
The notation $(a, b)\in{\leq}$ is used interchangeable with $a \leq b$ and $b \geq a$.
We call a pair of elements $a, b \in X$ \emph{comparable} if $a \leq b$ or $b \leq a$, otherwise we call them \emph{incomparable}.
A subset of $X$ where all elements are pairwise comparable is called a \emph{chain}.
An element $a$ is called \emph{strictly less than} an element $b$ if $a \leq b$ and $a \neq b$ and is denoted by $a < b$, the element $b$ is then called \emph{strictly greater than} $a$.
For an ordered set $(X,\leq)$, the associated covering relation ${\lessdot} \subset {<}$ is given by all pairs $(a,c)$ with $a < c$ for which no element $b$ with $a < b < c$ exists.
A \emph{graph} is a pair $(V,E)$ with $E \subset \binom{V}{2}$.
The set $V$ is called the set of \emph{vertices} and the set $E$ is called the set of \emph{edges}, two vertices $a$ and $b$ are called  \emph{adjacent} if $\{a,b\}\in E$.

From here out we give some notations in a way that is not necessarily standard but will be used throughout our work.
A $d$-dimensional \emph{order diagram} or \emph{drawing} of an ordered set $(X,\leq)$ is denoted by $(\vec{p}_a)_{a\in X}\subset \mathbb{R}^d$  whereby $\vec{p}_a=(x_{a,1},\ldots,x_{a,d-1},y_a)$ for each $a \in X$ and for all $a \lessdot b$ it holds that $y_a<y_b$.
Similarly, a $d$-dimensional \emph{graph drawing} of a graph $(V,E)$ is denoted by $(\vec{p}_a)_{a\in V}$  with $\vec{p}_a=(x_{a,1},\ldots,x_{a,d-1},y_a)$ for each $a \in V$.
If the dimension of a order diagram or a graph drawing is not qualified, the two-dimensional case is assumed.
In this case an order diagram can be depicted in the plane by visualizing the elements as a dot or \emph{nodes} and connecting element pairs in the covering relation by a straight line.
In the case of a graph, vertices are depicted by a dot and adjacent vertices are connected by a straight line.
We call $y_a$ the \emph{vertical component} and $x_{a,1},\ldots,x_{a,d-1}$ the \emph{horizontal components} of of $\vec{p}_a$ and denote $(\vec{p}_a)_x=(x_{a,1},\ldots,x_{a,d-1},0)$.
The forces operating on the vertical component are called the \emph{vertical force} and the forces operating on the horizontal components the \emph{horizontal forces}.
The Euclidean distance between the representation of $a$ and $b$ is denoted by $d(\vec{p}_a,\vec{p}_b) = |\vec{v}_a-\vec{v}_b|$, while the distance between the vertical components is denoted by $d_y(\vec{p}_a,\vec{p}_b)$ and the distance in the horizontal components is denoted by $d_x(\vec{p}_a,\vec{p}_b)=d((\vec{p}_a)_x,(\vec{p}_b)_x)$.
The unit vector from $\vec{p}_a$ to $\vec{p}_b$ is denoted by $\vec{u}(\vec{p}_a,\vec{p}_b)$, the unit vector operating in the horizontal dimensions is denoted by $\vec{u}_x(\vec{p}_a,\vec{p}_b)$.
Finally, the cosine-distance between two vector pairs $(\vec{a},\vec{b})$ and $(\vec{c},\vec{d})$ with $\vec{a},\vec{b},\vec{c},\vec{d}\in \mathbb{R}^d$ is given by $d_{\text{cos}}((\vec{a},{b}),(\vec{c},\vec{d}))\coloneqq 1-\frac{\sum_{i=1}^d(b_i-a_i)\cdot(d_i-c_i)}{d(a,b)\cdot d(c,d)}$.

\subsection{Force-Directed Graph Drawing}
\label{sec:force-direct-algor}

\begin{algorithm}[b]
  \caption{Force-Directed Algorithm by Eades}
  \label{alg:general}
  \begin{tabular}{llll}
    \textbf{Input:}  & Graph: $(V,E)$ & \textbf{Constants:} & $K\in \mathbb{N}$, $\varepsilon > 0$, $ \delta > 0$\\
                     & Initial drawing: $p =(\vec{p}_a)_{a \in V} \subset \R^2$ && \\
    \textbf{Output:} & Drawing: $p =(\vec{p}_a)_{a \in V} \subset \R^2$ &&
  \end{tabular}
  \hrule
\begin{lstlisting}
$t = 1$
while $t < K$ and $\max_{a\in V}\|F_a(t)\|>\varepsilon$:
  for $a \in V$:
    $F_a(t)\coloneqq \sum_{\{a,b\}\not\in E} f_{\text{rep}}(\vec{p}_a,\vec{p}_b) + \sum_{\{a,b\}\in E} f_{\text{spring}}(\vec{p}_a,\vec{p}_b)$
  for $a \in V$:
    $\vec{p}_a \coloneqq \vec{p}_a+\delta\cdot F_a(t)$
  $t = t + 1$
\end{lstlisting}
\end{algorithm}

The general idea of force-directed algorithms is to represent the graph as a physical model consisting of steel rings each representing a vertex.
For every pair of adjacent vertices, their respective rings are connected by identical springs.
Using a physical simulation, this system is then moved into a state of minimal stress, which can in turn be used as the drawing.
Many modifications to this general approach, that are not necessarily based on springs, were proposed
in order to encourage additional conditions in the resulting drawings.

\begin{figure}[t]
  \begin{minipage}{.32\linewidth}
    \includegraphics[width=\linewidth]{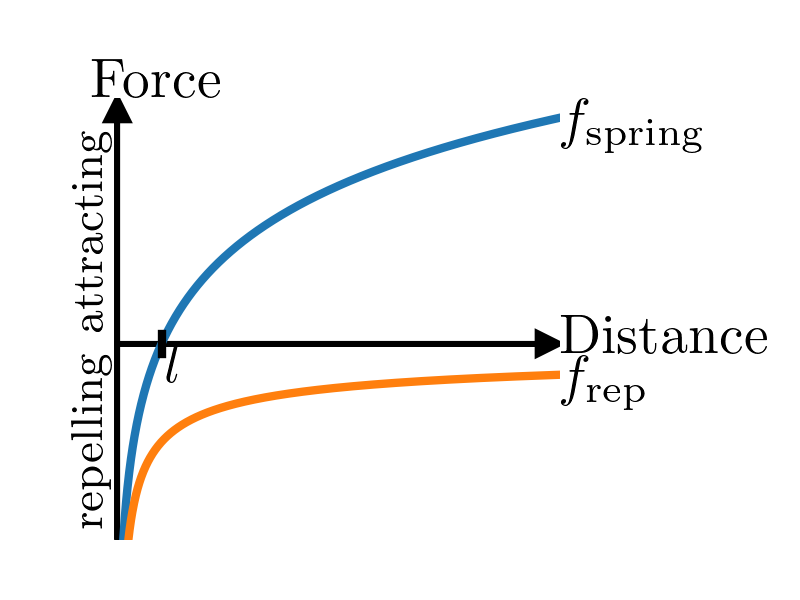}
    \caption{The forces for graphs as introduced by Eades in 1984. The $f_{\text{spring}}$ force operates between adjacent vertices and has an equilibrium at $l$, the force $f_{\text{rep}}$ is always a repelling force and operates on non-adjacent pairs.}
    \label{fig:eades}
  \end{minipage}
  \hfill
  \begin{minipage}{.32\linewidth}
    \includegraphics[width=\linewidth]{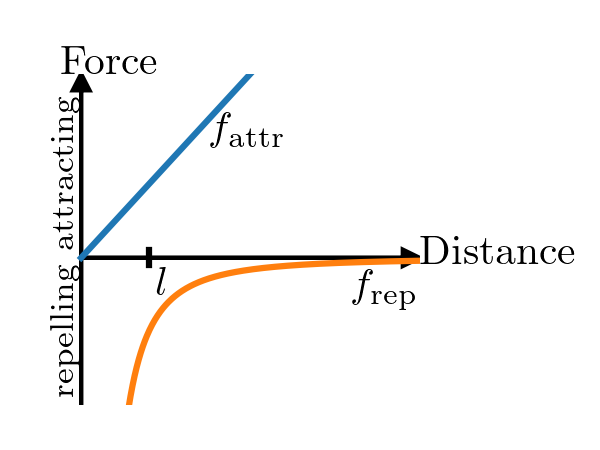}
    \caption{Horizontal forces for drawing order diagrams introduced by Freese in 2004. The force $f_{\text{attr}}$ operates between comparable pairs, the force $f_{\text{rep}}$ between incomparable pairs. There is no vertical force.}
    \label{fig:freese}
  \end{minipage}
  \hfill
  \begin{minipage}{.32\linewidth}
    \includegraphics[width=\linewidth]{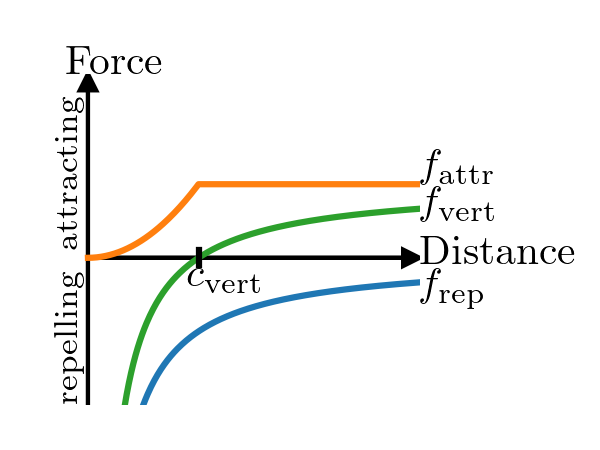}
    \caption{Our forces for drawing order diagrams. $f_{\text{vert}}$ operates vertically between node pairs in the covering relation, the force $f_{\text{attr}}$ between comparable pairs and the force $f_{\text{rep}}$ between incomparable pairs.}
    \label{fig:force}
  \end{minipage}
\end{figure}

The idea of force-directed algorithms was first suggested by Eades~\cite{Eades.1984}.
His algorithmic realization of this principle is done using an iterative approach where in each step of the simulation the forces that operate on each vertex are computed and summed up (cf.\ \cref{alg:general}).
Based on the sum of the forces operating on each vertex, they are then moved.
This is repeated for either a limited number of rounds or until there is no stress left in the physical model.
While a system consisting of realistic springs would result in linear forces between the vertices, Eades claims that those are performing poorly and thus introduces an artificial spring force.
This force operates on each vertex $a$ for adjacent pairs $\{a,b\}\in E$ and is given as
$  f_{\text{spring}}(\vec{p}_a,\vec{p}_b)=-c_{\text{spring}}\cdot \log\left(\frac{d(\vec{p}_a,\vec{p}_b)}{l}\right)\cdot \vec{u}(\vec{p}_a,\vec{p}_b)$,
whereby $c_{\text{spring}}$ is the spring constant and $l$ is the equilibrium length of the spring.
The spring force repels two vertices if they are closer then this optimal distance $l$ while it operates as an attracting force if two vertices have a distance greater then $l$, see \cref{fig:eades}.
To enforce that non-connected vertices are not placed too close to each other, he additionally introduces the repelling force that operates between non-adjacent vertex pairs as $f_{\text{rep}}(\vec{p}_a,\vec{p}_b)=\frac{c_{\text{rep}}}{d(\vec{p}_a,\vec{p}_b)^2}\cdot \vec{u}(\vec{p}_a,\vec{p}_b)$.
The value for $c_{\text{rep}}$ is once again constant.
In a realistic system, even a slightest movement of a vertex changes the forces that are applied to its respective ring.
To depict this realistically a damping factor $\delta$ is introduced in order to approximate the realistic system.
The smaller this damping factor is chosen, the closer the system is to a real physical system.
However, a smaller damping factor results in higher computational costs.
In some instances this damping factor is replaced by a cooling function $\delta(t)$ to guarantee convergence.
The physical simulation stops if the total stress of the system falls below a constant $\varepsilon$.
Building on this approach, a modification is proposed in the work of Fruchterman and Reingold \cite{Fruchterman.1991} from 1991.
In their algorithm, the force $f_{\text{attr}}(\vec{p}_a,\vec{p}_b)=-\frac{d(\vec{p}_a,\vec{p}_b)^2}{l} \cdot \vec{u}(\vec{p}_a,\vec{p}_b)$ is operating between every pair of connected vertices.
Compared to the spring-force in Eades' approach, this force is always an attracting force.
Additionally the force $  f_{\text{rep}}(\vec{p}_a,\vec{p}_b)=\frac{l^2}{d(\vec{p}_a,\vec{p}_b)}\cdot \vec{u}(\vec{p}_a,\vec{p}_b)$
repels every vertex pair.
Thus, the resulting force that is operating on adjacent vertices is given by $f_{\text{spring}}(\vec{p}_a,\vec{p}_b)=f_{\text{attr}}(\vec{p}_a,\vec{p}_b)+f_{\text{rep}}(\vec{p}_a,\vec{p}_b)$ and has once again its equilibrium at length $l$.
These forces are commonly considered to achieve better drawings than Eades' approach and are thus usually preferred.

While the graph drawing algorithms described above lead to sufficient results for undirected graphs, they are not suited for order diagram drawings as they do not take the direction of an edge into consideration.
Therefore, they will not satisfy the hard condition that greater elements have a higher $y$-coordinate.
Freese \cite{Freese.2004} thus proposed an algorithm for lattice drawing that operates in three dimensions, where the ranking function $rank(a) = height(a) - depth(a)$ fixes the vertical component.
The function $height(a)$ thereby evaluates to the length of the longest chain between $a$ and the minimal element and the function $depth(a)$ to the length of the longest chain to the maximal element.
While this ranking function guarantees that lesser elements are always positioned below greater elements, the horizontal coordinates are computed using a force-directed approach.
Freese introduces an attracting force between comparable elements that is given by $f_{\text{attr}}(\vec{p}_a,\vec{p}_b) = - c_{\text{attr}} \cdot d_x(\vec{p}_a,\vec{p}_b)\cdot \vec{u}_x(\vec{p}_a,\vec{p}_b)$,
and a repelling force that is given by $f_{\text{rep}}(\vec{p}_a,\vec{p}_b) = c_{\text{rep}} \cdot \frac{d_x(\vec{p}_a,\vec{p}_b)}{|y_b-y_a|^3+|x_{b,1}-x_{a,1}|^3+|x_{b,2}-x_{a,2}|^3}\cdot \vec{u}_x(\vec{p}_a,\vec{p}_b)$ operating on incomparable pairs only, (cf.\ \cref{fig:freese}).
The values for $c_{\text{attr}}$ and $c_{\text{rep}}$ are constants.
A parallel projection is either done by hand or chosen automatically in order to compute a two-dimensional depiction of the three-dimensional drawing.

\section{The ReDraw Algorithm}
\label{sec:algorithm}

Our algorithm \ReDraw{} uses a force-directed approach similar to the one that is used in Freese's approach.
Compared to Freese's algorithm, we however do not  use a static ranking function to compute the vertical positions  in the drawing.
Instead, we use forces which allow us to incorporate additional properties like the horizontal distance of vertex pairs, into the vertical distance.
By respecting a vertical invariant, that we will describe later, the vertical movement of the vertices is  restricted so that the hard constraint on the $y$-coordinates of comparable nodes can be always guaranteed.
However, the algorithm is thus more likely to get stuck in a local minimum.
We address this problem by computing the first drawing in a high dimension and then iteratively reducing the dimension of this drawing until a two-dimensional drawing is achieved.
As additional degrees of freedom allow the drawing to move less restricted in higher dimensions it thus reduces the probability for the system to get stuck in a local minimum.

\begin{algorithm}[b]
  \caption{ReDraw Algorithm}
  \label{alg:our}
  \begin{tabular}{llll}
    \textbf{Input:}  & Ordered set: $O=(X,\leq)$ &\textbf{Constants:} & $K\in \mathbb{N}$, $\varepsilon > 0$, $ \delta > 0$, \\
                     & Initial dimension: $d$&&$c_{\text{vert}}>0$, $c_{\text{hor}}>0$,\\
    \textbf{Output:} & Drawing: $p =(\vec{p}_a)_{a \in V} \subset \R^2$ &&$c_{\text{par}}>0$, $c_{\text{ang}}>0$, $c_{\text{dist}}>0$
  \end{tabular}
  \hrule
\begin{lstlisting}
$p$ = initial_drawing($O$)
while $d \geq 2$:
  node_step($O,p,d,K,\varepsilon,\delta,c_{\text{vert}},c_{\text{hor}}$)
  line_step($O,p,d,K,\varepsilon,\delta,c_{\text{par}},c_{\text{ang}},c_{\text{dist}}$)
  if $d > 2$:
    dimension_reduction($O,p,d$)
    $d$ = $d-1$

\end{lstlisting}
\end{algorithm}
Our algorithm framework (cf.\ \cref{alg:our}) consists of three individual algorithmic steps that are iteratively repeated.
We call one repetition of all three steps a \emph{cycle}.
In each cycle the algorithm is initialized with the $d$-dimensional drawing and returns a $(d-1)$-dimensional drawing.
The first step of the cycle, which we refer to as the \emph{node step}, improves the $d$-dimensional drawing by optimizing the proximity of nodes in order to achieve a better representation of the ordered set.
In the second step, which we call the \emph{line step}, the force-directed approach is applied to improve distances between different lines as well as between lines and nodes.
The resulting drawing thereby achieves a better satisfaction of soft criteria and thus improve the readability for a human reader.
Finally, in the \emph{reduction step} the  dimension of the drawing is reduced to $(d-1)$ by using a parallel projection into a subspace that preserves the vertical dimension.
In the last (two-dimensional) cycle, the dimension reduction step is omitted.

The initial drawing used in the first cycle is randomly generated.
The vertical coordinate of each element dot is given by its position in a randomly chosen linear extension of the ordered set.
The horizontal coordinates of each element are set to a random value between -1 and 1.
This guarantees that the algorithm does not start in an unstable local minimum.
Every further cycle then uses the output of the previous cycle as input to further enhance the resulting drawing.

Compared to the approach taken by Freese we do not fix the vertical component by a ranking function.
Instead, we recompute the vertical position of each element in each step using our force-directed approach.
To ensure that the resulting drawing is in fact a drawing of the ordered set we guarantee that in every step of the algorithm the following property is satisfied:
\begin{definition}
  Let $(X,\leq)$ be an ordered set with a drawing $(\vec{p}_a)_{a \in X}$.
  The drawing $(\vec{p}_a)_{a \in X}$ \emph{satisfies the vertical constraint}, iff. $\forall a,b\in X: a < b \Rightarrow y_a < y_b$.
\end{definition}
This vertical invariant is preserved in each step of the algorithm and thus that in the final drawing the comparabilities of the order are correctly depicted.

\subsection{Node Step}
\label{sec:force-step}

The first step of the iteration is called the \emph{node step}, which is used in order to compute a $d$-dimensional representation of the ordered set.
It thereby emphasizes the ordinal structure by positioning element pairs in a similar horizontal position, if they are comparable.
In this step we define three different forces that operate simultaneously.
For each $a \leq b$ on $a$ the vertical force
$f_{\text{vert}}(\vec{p}_a,\vec{p}_b)= \left(0,\ldots,0,- c_{\text{vert}} \cdot \left(\frac{1+d_x(\vec{p}_a,\vec{p}_b)}{d_y(\vec{p}_a,\vec{p}_b)}-1\right)\right)$
operates  while on $b$ the force $-f_{\text{vert}}(\vec{p}_a,\vec{p}_b)$ operates.
If two elements have the same horizontal coordinates it has its equilibrium if the vertical distance is at the constant $c_{\text{vert}}$.
Then, if two elements are closer then this constant it operates repelling and if they are farther away the force operates as an attracting force.
Thus, the constant $c_{\text{vert}}$ is a parameter that can be used to tune the \emph{optimal vertical distance}.
By incorporating the horizontal distance into the force, it can be achieved that vertices with a high horizontal distance will also result in a higher vertical distance.
Note, that this force only operates on the covering relation instead of all comparable pairs, as
Otherwise, chains would be contracted to be positioned close to a single point.

\begin{algorithm}[b]
  \caption{ReDraw - Node step}
  \label{alg:force_step}
  \begin{tabular}{llll}
    \textbf{Input:}  & Ordered set: $(X,\leq)$ & \textbf{Constants:} & $K\in \mathbb{N}$, $\varepsilon > 0$, $ \delta > 0$,\\
                     & Drawing $p =(\vec{p}_a)_{a \in X} \subset \R^d$ && $c_{\text{vert}}>0$, $c_{\text{hor}}>0$ \\
    \textbf{Output:} & Drawing: $p =(\vec{p}_a)_{a \in X} \subset \R^d$ &&
  \end{tabular}
  \hrule
\begin{lstlisting}
$t = 1$
while $t < K$ and $\max_{a\in X}\|F_a(t)\|>\varepsilon$:
  for $a \in X$:
    $F_a(t)\coloneqq \sum_{a \lessdot b} f_{\text{vert}}(\vec{p}_a,\vec{p}_b)-\sum_{b \lessdot a} f_{\text{vert}}(\vec{p}_a,\vec{p}_b)$
    $\phantom{F_a(t)\coloneqq}+ \sum_{a \leq b} f_{\text{attr}}(\vec{p}_a,\vec{p}_b) +  \sum_{a \not\leq b} f_{\text{rep}}(\vec{p}_a,\vec{p}_b)$
  for $a \in X$:
    $\vec{p}_a \coloneqq$ overshooting_protection($\vec{p}_a+\delta\cdot F_a(t)$)
  $t = t + 1$
\end{lstlisting}
\end{algorithm}

On the other hand there are two different forces that operate in horizontal direction.
Similar to Freese's layout, there is an attracting force between comparable and a repelling force between incomparable element pairs; however, the exact forces are different.
Between all comparable pairs $a$ and $b$ the force
 $f_{\text{attr}}(\vec{p}_a,\vec{p}_b) = - \min\left(d_x(\vec{p}_a,\vec{p}_b)^3,c_{\text{hor}}\right)\cdot \vec{u}_x(\vec{p}_a,\vec{p}_b)$
is operating.
Note that in contrast to $f_{\text{vert}}$ this force operates not only on the covering but on all comparable pairs and thus encourages chains to be drawn in a single line.
Similarly, incomparable elements should not be close to each other and thus the force
  $f_{\text{rep}}(\vec{p}_a,\vec{p}_b) =  \frac{c_{\text{hor}}}{d_x(\vec{p}_a,\vec{p}_b)} \cdot \vec{u}_x(\vec{p}_a,\vec{p}_b)$,
repels incomparable pairs horizontally.

We call the case that an element would be placed above a comparable greater element or below a lesser element, \emph{overshooting}.
However, to ensure that every intermediate drawing that is computed in the node step still satisfies the vertical invariant we have to prohibit overshooting.
Therefore, we add overshooting protection to the step in the algorithm where $(\vec{p}_a)_{a \in X}$ is recomputed.
This is done by restricting the movement of every element such that it is placed maximally  $\frac{c_{\text{vert}}}{10}$ below the lowest positioned greater element, or symmetrically above the greatest lower element.
If the damping factor is chosen sufficiently small overshooting is rarely required.
This is, because our forces are defined such that the closer two elements are positioned the stronger they repel each other, see \cref{fig:force}.

All three forces are then consolidated into a single routine that is repeated at most $K$ times or until the total stress falls below a constant $\varepsilon$, see \cref{alg:force_step}.
The general idea of our forces is similar to the forces described in Freese's approach, as comparable elements attract each other and incomparable elements repel each other.
However, we are able to get rid of the ranking function that fixes $y$-coordinate and thus have an additional degree of freedom which allows us to include the horizontal distance as a factor to determine the vertical positions.
Furthermore, our forces are formulated in a general way such that the drawings can be computed in arbitrary dimensions, while Freese is restricted to three dimensions.
This overcomes the problem of getting stuck in local minima and enables us to recompute the drawing in two dimensions in the last cycle.

\subsection{Line Step}
\label{sec:parralellize-step}

\begin{algorithm}[t]
  \caption{ReDraw - Line Step}
  \label{alg:parallelize_step}
  \begin{tabular}{llll}
    \textbf{Input:}  & Ordered set: $(X,\leq)$ & \textbf{Constants:} & $K\in \mathbb{N}$, $\varepsilon > 0$, $ \delta > 0$,\\
                     & Drawing $p =(\vec{p}_a)_{a \in X} \subset \R^d$ && $c_{\text{par}}>0$, $c_{\text{ang}}>0$, $c_{\text{dist}}>0$ \\
    \textbf{Output:} & Drawing: $p =(\vec{p}_a)_{a \in X} \subset \R^d$ &&
  \end{tabular}
  \hrule
\begin{lstlisting}
$t = 1$
while $t < K$ and $\max_{a\in X}\|F_a(t)\|>\varepsilon$:
  $A = \{\{(a,b),(c,d)\}\mid a\lessdot b,c \lessdot d, d_{\text{cos}}((\vec{p}_a,\vec{p}_b),(\vec{p}_c,\vec{p}_d)) < c_{\text{par}}\}$
  $B = \{\{(a,c),(b,c)\}\mid (a\lessdot c, b\lessdot c) \text{ or }  (c \lessdot a, c\lessdot b), d_{\text{cos}}((\vec{p}_a,\vec{p}_c),(\vec{p}_b,\vec{p}_c)) < c_{\text{ang}}\}$
  $C = \{(a,(b,c))\mid a \in X, b\lessdot c,  d(\vec{p}_a,(\vec{p}_b,\vec{p}_c))< c_{\text{dist}}\}$
  for $a \in X$:
    $F_a(t)\coloneqq \sum_{\{(a,b),(c,d)\}\in A} f_{\text{par}}((\vec{p}_a,\vec{p}_b),(\vec{p}_c,\vec{p}_d))+ \sum_{(a,(b,c)) \in C} f_{\text{dist}}(\vec{p}_a,(\vec{p}_b,\vec{p}_c)) $
    $\phantom{F_a(t)}- \sum_{\{(b,a),(c,d)\}\in A} f_{\text{par}}((\vec{p}_a,\vec{p}_b),(\vec{p}_c,\vec{p}_d)) - \frac{1}{2} \sum_{(b,(a,c)) \in C} f_{\text{dist}}(\vec{p}_a,(\vec{p}_b,\vec{p}_c)) $
    $\phantom{F_a(t)} + \sum_{\{(a,c),(b,c)\}\in B} f_{\text{ang}}((\vec{p}_a,\vec{p}_c),(\vec{p}_b,\vec{p}_c)) $
  for $a \in X$:
    $\vec{p}_a \coloneqq$ overshooting_protection($\vec{p}_a+\delta\cdot F_a(t)$)
  $t = t + 1$
\end{lstlisting}
\end{algorithm}
While the goal of the node step is to get a good representation of the internal structure by optimizing on the proximity of nodes, the goal of the line step is to make the resulting drawing more aesthetically pleasing by optimizing distances between lines.
Thus, in this step the drawing is optimized on three soft criteria.
First, we want to maximize the number of parallel lines.
Secondly, we want to achieve large angles between two lines that are connected to the same element dot.
Finally, we want to have a high distance between elements and non-adjacent lines.
We achieve a better fit to these criteria by applying a force-directed algorithm with three different forces, each optimizing on one criterion.
While the previous step does not directly incorporate the path of the lines, this step incorporates those into its forces.
Therefore, we call this step the \emph{line step}.

The first force of the line step operates on lines $(a,b)$ and $(c,d)$ with $a\neq c$ and $b \neq d$   if their  cosine distance is below a threshold $c_{\text{par}}$.
The horizontal force
$  f_{\text{par}}((\vec{p}_a,\vec{p}_b),(\vec{p}_c,\vec{p}_d)) = -\left(1-\frac{d_{\text{cos}}((\vec{p}_a,\vec{p}_b),(\vec{p}_c,\vec{p}_d))}{c_{\text{par}}}\right)\cdot\left(\frac{(\vec{p}_b-\vec{p}_a)_x}{y_b-y_a} - \frac{(\vec{p}_d-\vec{p}_c)_x}{y_d-y_c}\right)$
operates on $a$ and the force $-f_{\text{par}}((\vec{p}_a,\vec{p}_b),(\vec{p}_c,\vec{p}_d))$ operates to $b$.
This result of this force is thus that almost parallel lines are moved to become more parallel.
Note, that this force becomes stronger the more parallel the two lines are.

The second force operates on lines that are connected to the same dot and have a small angle, \ie, lines with cosine distance below a threshold $c_{\text{ang}}$.

Let $(a,c)$ and $(b,c)$ be such a pair then the horizontal force operating on $a$ is given by
$f_{\text{ang}} ((\vec{p}_a,\vec{p}_c),(\vec{p}_b,\vec{p}_c)) =  \left(1-\frac{d_{\text{cos}}((\vec{p}_a,\vec{p}_c),(\vec{p}_b,\vec{p}_c))}{c_{\text{ang}}}\right)\cdot\left(\frac{(\vec{p}_c-\vec{p}_a)_x}{y_c-y_a} - \frac{(\vec{p}_c-\vec{p}_b)_x}{y_c-y_b}\right).$
In this case, once again the force is stronger for smaller angles; however, the force is operating in the opposite direction compared to $f_{\text{par}}$ and thus makes the two lines less parallel.
Symmetrically, for each pair $(c,a)$ and $(c,b)$ the same force operates on $a$.
There are artifacts from $f_{\text{par}}$ that operate against $f_{\text{ang}}$ in opposite direction.
This effect should be compensated for by using a much higher threshold constant $c_{\text{ang}}$ than $c_{\text{par}}$, otherwise the benefits of this force are diminishing.

Finally, there is a force that operates on all pairs of element dots $a$ and lines $(b,c)$, for which the distance between the element and the line is closer then  $c_{\text{dist}}$.
The force
$f_{\text{dist}} (\vec{p}_a,(\vec{p}_b,\vec{p}_c)) = \frac{1}{d(\vec{p}_a,(\vec{p}_b,\vec{p}_c))}\cdot\left(  (\vec{p}_a-\vec{p}_c)-\frac{(\vec{p}_a-\vec{p}_c)\cdot(\vec{p}_b-\vec{p}_c)}{(\vec{p}_b-\vec{p}_c)\cdot (\vec{p}_b-\vec{p}_c)}(\vec{p}_b-\vec{p}_c)\right)$
is applied to $a$ and $-f_{\text{dist}}(a,(c,d))/2$ is applied to $b$ and $c$.
This results in a force whose strength is linearly stronger, the closer the distance $d(\vec{p}_a,(\vec{p}_b,\vec{p}_c))$.
It operates in perpendicular direction to the line and repels the dot and the line.

Similar to the node step, all three forces are combined into a routine that is repeated until the remaining energy in the physical system drops below a certain stress level $\varepsilon$.
Furthermore a maximal number of repetitions $K$ is fixed.
We also once again include the overshooting protection as described in the previous section to make sure that the vertical invariant stays satisfied.

The line step that is described in this section is a computational demanding task, as in every repetition of the iterative loop the sets of almost parallel edges, small angles and elements that are close to lines have to be recomputed.
To circumvent this problem on weaker hardware, there are a number of possible speedup techniques.
First of all, the sets described above do not have to be recomputed every iteration, but can be cashed over a small number of iterations.
In \cref{alg:parallelize_step} these are the sets $A$, $B$ and $C$.
By recomputing those sets only every $k$-th iteration a speedup to almost factor $k$ can be achieved.
Another speedup technique that is possible is to only execute the line step in the last round.
Both of these techniques however have a trade off for the quality of the final drawing and are thus not further examined in this paper.

\subsection{Dimension Reduction}
\label{sec:stepdown}

In the dimension reduction step, we compute a $(d-1)$-dimensional drawing from the $d$-dimensional drawing with the goal of reflecting the structural details of the original drawing like proximity and angles.
Our approach to solve this is to compute a $(d-1)$-dimensional linear subspace of the $d$-dimensional space.
By preserving the vertical dimension we can ensure that the vertical invariant stays satisfied.
Then a parallel projection into this subspace is performed.

As such a linear subspace always contains the origin, we center our drawing around the origin.
Thereby, the whole drawing $(\vec{p}_a)_{a \in X}$ is geometrically translated such that the mean of every coordinate becomes 0.
The linear subspace projection is performed as follows:
The last coordinate of the linear subspace will be the vertical component of the $d$-dimensional drawing to ensure that the vertical invariant is preserved.
For the other $(d-1)$ dimensions of the original space, a principle component analysis~\cite{Pearson.1901} is performed to reduce them to a $(d-2)$-dimensional subspace.
By combining this projection with the vertical dimension a $(d-1)$-dimensional drawing is achieved, that captures the structure of the original, higher-dimensional drawing and represents its structural properties.

It is easily possible to replace PCA in this step by any other dimension reduction technique.
It would thus be thinkable to just remove the first coordinate in each step and hope that the drawing in the resulting subspace has enough information encapsulated in the remaining coordinates.
Also other ways of choosing the subspace in which is projected could be considered.
Furthermore, non-linear dimension reduction methods could be tried in order to achieve drawings, however our empirical experiments suggest, that PCA hits a sweet spot.
The payoff of more sophisticated dimension reduction methods seems to be negligible as each drawing is further improved in lower dimensions.
On the other hand we observed local minima if we used simpler dimension reduction methods.

\section{Evaluation}
\label{sec:evaluation}

As we described in the previous sections, it is not a trivial task to evaluate the quality of an order diagram drawing.
Drawings that one human evaluator might consider as favorably might not be perceived as readable by others.
Therefore, we evaluate our generated drawing with a large quantity of domain experts.

\begin{figure}[t]
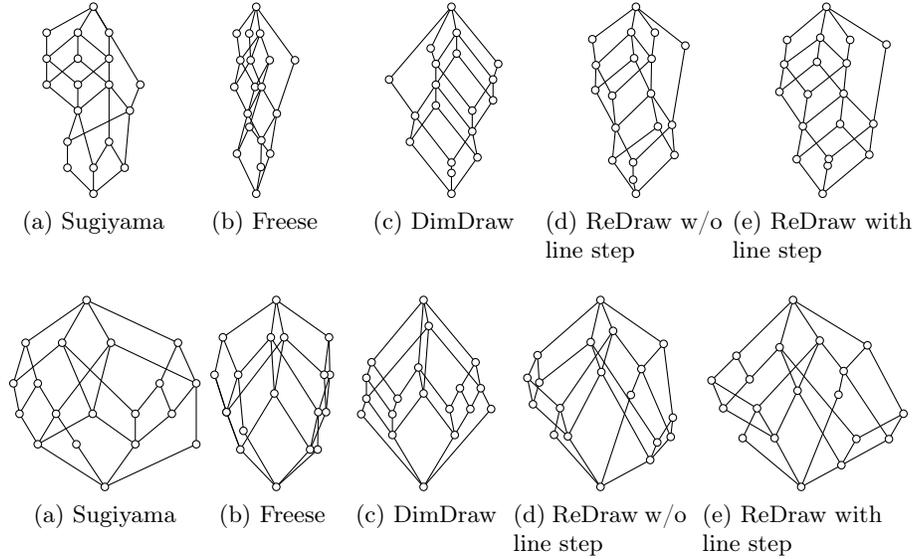

  \centering
\subfloat[Sugiyama]{\hspace*{1.75em}\includegraphics[height=8em]{tikz/forum_sugi.tikz}\hspace*{1.75em}}
  \subfloat[Freese]{\hspace*{1.75em}\includegraphics[height=8em]{tikz/forum_freese.tikz}\hspace*{1.75em}}
  \subfloat[DimDraw]{\hspace*{1.75em}\includegraphics[height=8em]{tikz/forum_dimdraw.tikz}\hspace*{1.75em}}
  \subfloat[ReDraw w/o \\ line step]{\hspace*{1.75em}\includegraphics[height=8em]{tikz/forum_before.tikz}\hspace*{1.75em}}
  \subfloat[ReDraw with \\ line step]{\hspace*{1.75em}\includegraphics[height=8em]{tikz/forum.tikz}\hspace*{1.75em}}
  \\
  \setcounter{subfigure}{0} \subfloat[Sugiyama]{\hspace*{0.2em}\includegraphics[height=8em]{tikz/fish_sugi.tikz}\hspace*{0.2em}}
\subfloat[Freese]{\hspace*{0.2em}\includegraphics[height=8em]{tikz/fish_freese.tikz}\hspace*{0.2em}}
  \hspace*{0.2em}
  \subfloat[DimDraw]{\hspace*{0.2em}\includegraphics[height=8em]{tikz/fish_dimdraw.tikz}\hspace*{0.2em}}
  \hspace*{0.2em}
  \subfloat[ReDraw w/o\\line step]{\hspace*{0.4em}\includegraphics[height=8em]{tikz/fish_before.tikz}\hspace*{0.4em}}
  \hspace*{0.3em}
  \subfloat[ReDraw with\\line step]{\hspace*{0.2em}\includegraphics[height=8em]{tikz/fish.tikz}\hspace*{0.2em}}
  \caption{Top: Drawing of the lattices for the formal contexts ``forum romanum'' (top) and ``living beings and water'' (bottom) from the test dataset.}
  \label{fig:forum}
\end{figure}

\begin{figure}[t]
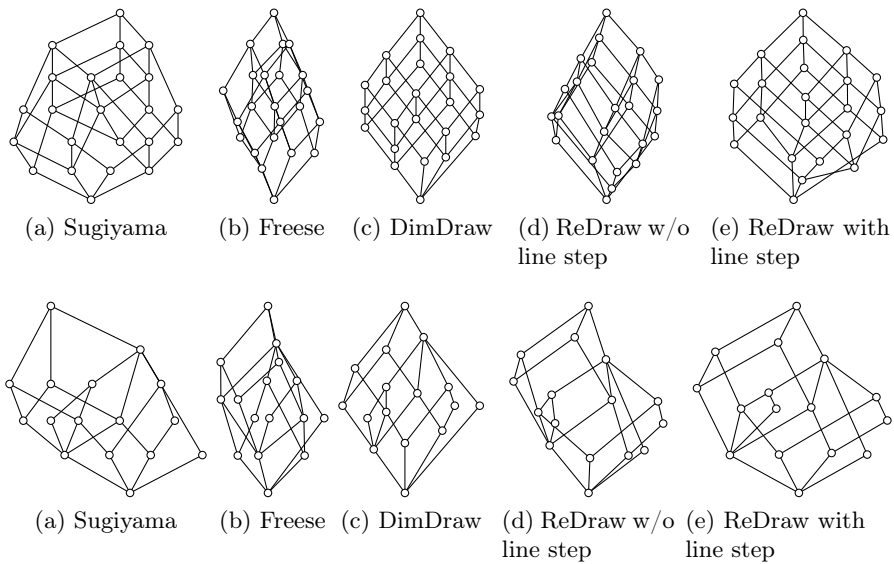

  \centering
\subfloat[Sugiyama]{\hspace*{.75em}\includegraphics[height=8em]{tikz/therapy_sugi.tikz}\hspace*{.75em}}
  \subfloat[Freese]{\hspace*{.75em}\includegraphics[height=8em]{tikz/therapy_freese.tikz}\hspace*{.75em}}
  \subfloat[DimDraw]{\hspace*{.75em}\includegraphics[height=8em]{tikz/therapy_dimdraw.tikz}\hspace*{.75em}}
    \hspace*{0.3em}
  \subfloat[ReDraw w/o\\line step]{\hspace*{1.2em}\includegraphics[height=8em]{tikz/therapy_before.tikz}\hspace*{1.1em}}
    \hspace*{0.5em}
    \subfloat[ReDraw with\\line step]{\hspace*{.75em}\includegraphics[height=8em]{tikz/therapy.tikz}\hspace*{.75em}}
    \\\setcounter{subfigure}{0}
    \subfloat[Sugiyama]{\hspace*{0.2em}\includegraphics[height=8em]{tikz/ice_sugi.tikz}\hspace*{0.2em}}
\subfloat[Freese]{\hspace*{0.2em}\includegraphics[height=8em]{tikz/ice_freese.tikz}\hspace*{0.2em}}
\subfloat[DimDraw]{\hspace*{0.2em}\includegraphics[height=8em]{tikz/ice_dimdraw.tikz}\hspace*{0.2em}}
\hspace*{0.2em}
\subfloat[ReDraw w/o\\ line step]{\hspace*{0.3em}\includegraphics[height=8em]{tikz/ice_before.tikz}\hspace*{0.3em}}
\hspace*{0.1em}
\subfloat[ReDraw with\\ line step]{\hspace*{0.3em}\includegraphics[height=8em]{tikz/ice.tikz}\hspace*{0.2em}}
  \caption{Top: Drawing of the lattices for the formal contexts ``therapy'' (top) and ``ice cream'' (bottom) from the test dataset.}
  \label{fig:therapy}
\end{figure}

\subsection{Run-Time Complexity}
\label{sec:empirical-evaluation}
The run-time of the node step is limited by $\mathcal{O}(n^2)$ with $n$ being the number of elements, as the distances between every element pair are computed.
The run-time of the edge step is limited by $\mathcal{O}(n^4)$, as the number of lines is bounded by $\mathcal{O}(n^2)$.
Finally, the run-time of the reduction step is determined by PCA which is known
to be bounded by $O(n^3)$.
Therefore, the total run-time of the algorithm is polynomial in $\mathcal{O}(n^4)$.
This is an advantage compared to DimDraw and Sugiyama's framework, which both solve exponential problems; however, Sugiyama is usually applied with a combination of heuristics to overcome this problem.
Freese's layout has by its nature of being a force-directed order diagram
drawing algorithm, similar to our approach, polynomial run-time.
Thus, for larger diagrams, only \ReDraw{}, Freese's algorithm and Sugiyama's framework (the latter with its heuristics) are suitable, while DimDraw is not.

\subsection{Tested Datasets}
Our test dataset consists of 77 different lattices including all classical examples of lattices described in~\cite{fcabook}.
We enriched these by lattices of randomly generated contexts and some sampled contexts from large binary datasets.
An overview of all related formal contexts for these lattices, together with their drawing generated by \ReDraw{} is published together with its source code.
We restrict the test dataset to lattices, as lattice drawings are of great interest for the formal concept analysis community.
This enables us to perform a user study using domain experts for lattices from the FCA community to evaluate the algorithm.

\subsection{Recommended Parametrizations}
As it is hardly possible to conduct a user study for every single combination of parameters, our  recommendations are based on empirical observations.
We used a maximal number of $K=1000$ algorithm iterations or stopped if the stress in the physical system fell below $\varepsilon=0.0025$.
Our recommended damping factor $\delta=0.001$.
In the node step we set $c_{\text{vert}}=1$ as the optimal horizontal distance and $c_{\text{hor}}=5$.
We used the thresholds $c_{\text{par}}=0.005$, $c_{\text{ang}}=0.05$ and $c_{\text{dist}}=1$ in the line step.
The drawing algorithms are started with 5 dimensions as we did not observe any notable improvements with higher dimensional drawings.
Finally the resulting drawing is scaled in horizontal direction by a factor of 0.5.

\subsection{Empirical Evaluation}

To demonstrate the quality of our approach we compare the resulting drawings to the drawings generated by a selected number of different algorithms in \cref{fig:forum} and \cref{fig:therapy}.
The different drawings are computed using Sugiyama's framework, Freese's layout, DimDraw and our new approach.
Additionally, a drawing of our approach before the line step is presented to show the impact of this line step.
In the opinion of the authors of this paper, the approach proposed in this paper achieves satisfying results for these ordered sets.
In most cases, we still prefer the output of DimDraw (and sometimes Sugiyama), but \ReDraw{} is able to cope with much larger datasets because of its polynomial nature.
Modifications of \ReDraw{} that combine the node step and the edge step into a single step were tried by the authors; however, the then resulting algorithm did not produce the anticipated readability, as the node and edge forces seem to work against each other.

\subsection{User Evaluation}

To obtain a measurable evaluation we  conducted a user study to compare the different drawings generated by our algorithm to two other algorithms.
We decided to compare our approach to Freese's and Sugiyama's algorithm, as those two seem to be the two most popular algorithms for lattice drawing at the moment.
We decided against including DimDraw into this study as, even though it is known to produce well readable drawings, it struggles with the computational costs for drawings of higher order dimensions due to its exponential nature.

\textbf{Experimental Setup.}
In each step of the study, all users are presented with three different drawings of one lattice from the dataset in random order and have to decide which one they perceive as ``most readable''.
The term ``most readable'' was neither further explained nor restricted.

\begin{figure}[t]
  \centering
  \begin{minipage}{0.3\linewidth}
    \vspace*{1.5em}
    \includegraphics[width=.95\textwidth]{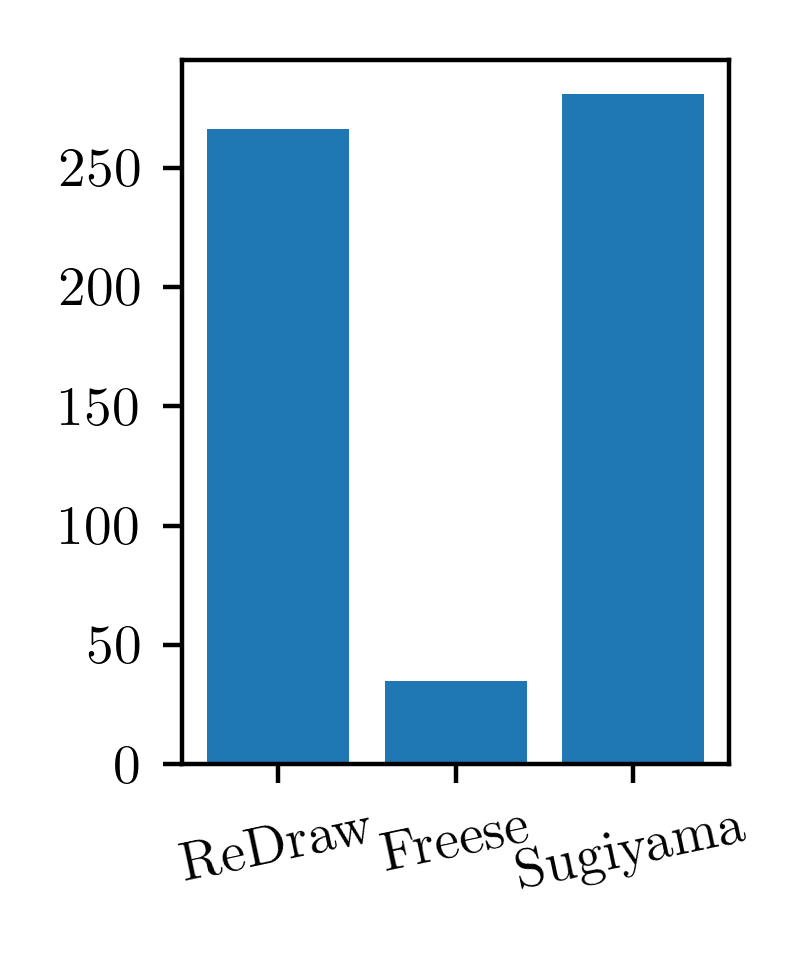}
  \end{minipage}
  \begin{minipage}{0.6\linewidth}
    \includegraphics[width=.95\textwidth]{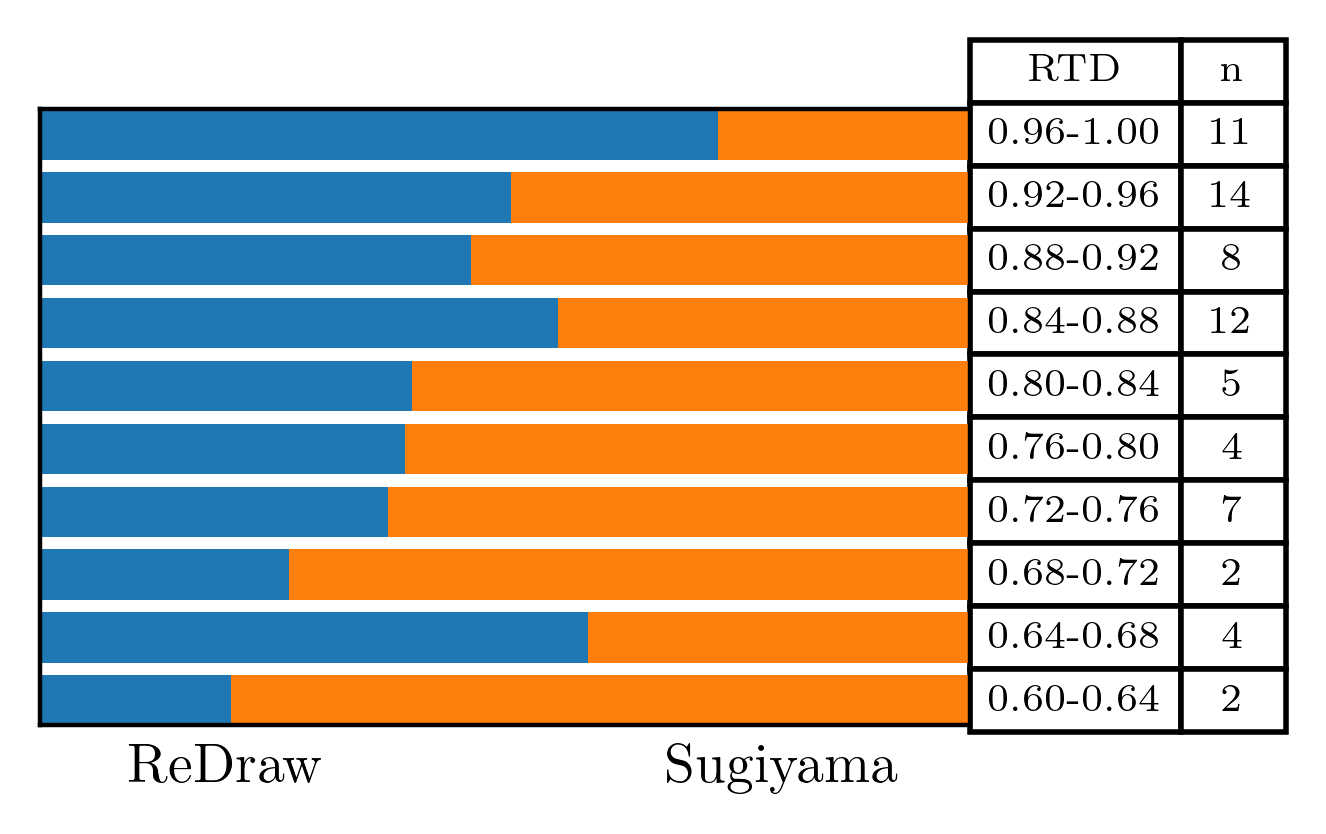}
  \end{minipage}
  \caption{Results of the user study. L: Number of votes for each algorithm. R: Share of votes for ordered sets divided into ranges of different truncated distributivity.}
  \label{fig:user}
\end{figure}

\textbf{Results.}
The study was conducted with nine experts from the formal concept analysis community to guarantee expertise with order diagrams among the participants.
Thus, all ordered sets in this study were lattices.
The experts voted 582 times in total; among those votes, 35 were cast for Freese's algorithm, 266 for our approach and 281 for Sugiyama.
As a common property of lattices is to contain a high degree of truncated distributivity~\cite{Wille.2003}, which makes this property of special interest,
we decided to compute the share of distributive triples for each lattice
excluding those resulting in the bottom-element.
We call the share of such distributive triples of all possible triples the
\emph{truncated relative distributivity (RTD)}.
Based on the RTD we compared the share of votes for Sugiyama's framework and \ReDraw{} for all order diagrams that are in a specific truncated distributivity range.
The results of this comparison are depicted in \cref{fig:user}.
The higher the RTD, the better \ReDraw{} performs in comparison.
The only exception in the range 0.64-0.68 can be traced back to a small test set with $n=4$.

\textbf{Discussion.}
As one can conclude from the user study, our force-directed algorithm performs on a similar level to Sugiyama's framework while outperforming Freese's force-directed layout.
In the process of developing \ReDraw{} we also conducted a user-study that compared an early version to DimDraw which suggested that \ReDraw{} can't compete with DimDraw.
However, DimDraw's exponential run-time makes computing larger order drawings unfeasible.
From the comparison of \ReDraw{} and Sugiyama's, that takes the RTD into account, we can follow that our algorithm performs better on lattices the higher the RTD.
We observed similar results when we computed the relative normal distributivity.
The authors of this paper thus recommend to use \ReDraw{} for larger drawings that are highly distributive.
Furthermore, the authors observed, that \ReDraw{} performs better if there are repeating structures or symmetries in the lattice as each instance of such a repetition tends to be drawn similarly.
This makes it the algorithm of choice for ordered sets that are derived from datasets containing high degrees of symmetries.
Anyway, the authors of this paper are convinced that there is no single drawing algorithm that can produce readable drawings for all different kinds of order diagrams.
It is thus always recommended to use a combination of different algorithms and then decide on the best drawing.

\section{Conclusion and Outlook}
\label{sec:conclusion}

In this work we introduced our novel approach \ReDraw{} for drawing diagrams.
Thereby we adapted a force-directed algorithm to the realm of diagram drawing.
In order to guarantee that the emerging drawing satisfies the hard conditions of order diagrams we introduced a vertical invariant that was satisfied in every step of the algorithm.
The algorithm consists of two main ingredients, the first being the node step that optimizes the drawing in order to represent structural properties using the proximity of nodes.
The second is the edge step that improves the readability for a human reader by optimizing the distances of lines.
Of particular interest is the line step that enhances the quality of the produced drawings as, to our knowledge, it is the first of its kind.
To avoid local minima, our drawings are first computed in a high dimension and then iterativly reduced into two dimensions.
To make the algorithm easily accessible, we published the source code and gave recommendations for parameters.
Generated drawings were, in our opinion, suitable to be used for ordinal data analysis.
A study using domain experts to evaluate the quality of the drawings confirmed this observation.

Further work in the realm of order diagram drawing could be to modify the line step and combine it with algorithms such as  DimDraw.
Also modifications that produce additive drawings are of great interest and should be investigated further.
Finally, in the opinion of the authors the research fields of ordinal data analysis and graph drawing would benefit significantly from the establishment of a ``readability measure'' or at least of a decision procedure that, given two visualizations of the same ordered set identifies the more readable one.

\bibliographystyle{splncs04}
\bibliography{paper.bib}

\end{document}